\def\dend#1{{\if*#1{\it Paenibacillus dendritiformis}\else
                {\it P. dendritiformis}\fi}}
\def\Tvar{var. {\it dendron}}
\def\Cvar{var. {\it chiralis}}
\def\Tname#1{{\if*#1\dend* \Tvar\else
                \if-#1\dend{} \Tvar\else
                 \dend{} \Tvar{} #1\fi\fi}}
\def\Cname#1{{\if*#1\dend* \Cvar\else
                \if-#1\dend{} \Cvar\else
                 \dend{} \Cvar{} #1\fi\fi}}
\def\Vname#1{{\if*#1\eddi{Paenibacillus}{V}\else
                \if-#1\eddi{P.}{V}\else \eddi{P.}{V} #1\fi\fi}}
\def\bsub#1{{\if*#1{\it Bacillus subtilis}\else
                \if-#1{\it B. subtilis}\else {\it B. subtilis} #1\fi\fi}}
\def\bacil#1{\if *#1{Bacillus}\else{B.}\fi}
\def\bcirc#1{\if *#1{\it \bacil* circulans}\else
                \if -#1{\it \bacil{} circulans}\else
                   {\it \bacil{} circulans} #1\fi\fi}
\def\ecoli#1{{\if*#1{\it Escherichia coli}\else
                \if-#1{\it E. coli}\else {\it E. coli} #1\fi\fi}}
\def\salmon#1{{\if*#1{\it Salmonella typhimurium}\else
                \if-#1{\it S. typhimurium}\else
                {\it S. typhimurium} #1\fi\fi}}

\def\myxo#1{{\if*#1{\it Myxococcus xanthus}\else
                \if-#1{\it M. xanthus}\else
                {\it M. xanthus} #1\fi\fi}}

\def\partderiv#1#2{{\partial #1\over\partial #2}}

\def\T{{${\cal T }$} }

\def\Tme#1{{\T morphotype#1}}

\def\etal{{\it et al. }}

\def\diffusion{di\!f\!fusion}

\documentstyle[pre,aps,preprint,epsf]{revtex}
\input{psfig}
\begin{document}

\title{Lubricating Bacteria Model
  for Branching growth of Bacterial Colonies}
\author{Yonathan Kozlovsky, Inon Cohen, Ido Golding and Eshel Ben-Jacob}
\address{School of Physics and Astronomy,\\ Raymond and Beverly Sackler
  Faculty of Exact Sciences, \\ Tel Aviv University,
Tel Aviv 69 978, Israel}
\maketitle

\begin{abstract}
\baselineskip 18pt
Various bacterial strains (e.g. strains belonging to the genera
{\it Bacillus, Paenibacillus, Serratia} and
 {\it Salmonella}) exhibit colonial branching patterns during
growth on poor semi-solid substrates. These patterns reflect the bacterial
cooperative self-organization.
Central part of the cooperation is the collective formation of
 lubricant on top of the agar
which enables the bacteria to swim. 
Hence it provides the colony means to advance towards the food. 
One method of modeling the colonial development is via
coupled reaction-diffusion 
equations which describe the time evolution of the bacterial 
density and the 
concentrations of the relevant chemical fields. 
This idea has been pursued by a number of groups.
Here we present an additional model which specifically 
includes an evolution equation for the
lubricant excreted by the bacteria. We show that when the 
diffusion of the fluid is
governed by nonlinear diffusion coefficient branching patterns evolves.
We study the effect of the rates of emission and decomposition 
of the lubricant fluid on the 
observed patterns. The results are compared with experimental observations.
We also include fields of chemotactic agents and food chemotaxis 
and conclude 
that these features are needed in order to explain the observations.
\end{abstract}
\baselineskip 23pt

\newpage

\section{Introduction}

It is now understood that the study of cooperative self-organization
of bacterial colonies is an exciting new multidisciplinary field of
research, necessitating the merger of biological information with the
physics of non-equilibrium processes and the mathematics of
non-linear dynamics.
At this stage, several experimental systems have been identified, and
preliminary modeling efforts are making significant progress in
providing a framework for the understanding of experimental
observations \cite{MKNIHY92,MS96,Kessler85,FM89,PK92a,BSST92,MHM93%
,BSTCCV94a,BCSALT95,WTMMBB95,BCCVG97,KW97,ES98}.

In nature bacterial colonies must often cope with hostile 
environmental conditions.
To do so bacteria have developed sophisticated
cooperative behavior and intricate communication capabilities  
\cite{SDA57,Shap88,BenJacob97,BCL98,LB98}. These include: 
direct cell-cell physical interactions via extra-membrane
polymers \cite{Mend78,Devreotes89}, collective production of
extracellular "wetting" fluid for movement on hard surfaces
\cite{MKNIHY92,Harshey94}, long range chemical signaling, such as
quorum sensing \cite{FWG94,LWFBSLW95,FWG96} and 
chemotactic signaling\footnote{
  Chemotaxis is a bias of
movement according to the gradient of a chemical agent. Chemotactic
signaling is a chemotactic response to an agent emitted by the bacteria.}
\cite{BB91,BE95,BB95}, collective activation and deactivation of genes
\cite{ST91,SM93,MS96} and even exchange of genetic material
\cite{GR95,RPF95,Miller98}. Utilizing these capabilities, 
bacterial colonies
develop complex spatio-temporal patterns in response to adverse growth
conditions.

For researchers in the pattern formation field, the above 
communication mechanisms 
open a new class of tantalizing complex models exhibiting a much
richer spectrum of patterns than the models of non-living systems.

Fujikawa and Matsushita \cite{FM89,MF90,FM91} reported for the
first time \footnote{
We refer to the first time that branching growth was studied as
such. Observations of branching colonies occurred long ago
\protect\cite{SC38,Henrici48}.
} that bacterial colonies could grow elaborate branching patterns of
the type known from the study of fractal formation in the process
of diffusion-limited-aggregation (DLA) \cite{WitSan81,Sander86,Vicsek89} .
This work was done with \bsub*, but was subsequently extended to other
bacterial species such as {\it Serratia marcescens} and {\it
Salmonella anatum} \cite{MM95}.

Motivated by these observations, Ben-Jacob \etal 
conducted new experiments with a new species of bacteria that has been 
isolated from cultures of \bsub* \cite{BSST92,BTSA94,BSTCCV94a}.
The new species was designated \Tname* \cite{TBG98}.
This species is motile on the hard surface and its colonies
exhibit branching patterns (Fig. 1).
The new mode of tip-splitting growth was found to be inheritable and
transferable by a single cell, hence it is
referred to as a distinctive morphotype \cite{BCG98}, and, to indicate the
tip-splitting character of the growth, it was denoted \Tme.
In the next section we describe in some detail the observations of
Ben-Jacob \etal .
Additional studies of branching
colonial growth are reported by Matsuyama \etal \cite{MKNIHY92,MM93}
and Mendelson and Salhi \cite{MS96}.

How should one approach the modeling of the complex bacterial patterning?
With present computational power it is natural to use computer models as a
main tool in the study of complex systems. However, one must be careful not
to be trapped in the ``reminiscence syndrome'', described by J. D. Cowan
\cite{Horgan95}, as the tendency to devise a set of rules which will
mimic some aspect of the observed phenomena and then, to quote J. D.
Cowan "They say: `Look, isn't this reminiscent of a biological or
physical phenomenon!' They jump in right away as if it's a decent
model for the phenomenon, and usually of course it's just got some
accidental features that make it look like something." Yet the
reminiscence modeling approach has some indirect value.  True, doing
so does not reveal (directly) the biological functions and behavior.
However, it does reflect understanding of geometrical and temporal
features of the patterns, which indirectly might help in revealing
the underlying biological principles. Another extreme is the
"realistic modeling" approach, where one constructs an algorithm that
includes in details all the known biological facts about the system.
Such an approach sets a trajectory of ever including more and more
details (vs. generalized features). The model keeps evolving to
include so many details that it loses any predictive power.

Here we try to promote another approach -- the "generic modeling" one
\cite{KL93,BSTCCV94a,Azbel93,KW97}. We seek to elicit, from the
experimental observations and the biological knowledge, the generic
features and basic principles needed to explain the biological
behavior and to include these features in the model. We will
demonstrate that such modeling, with close comparison to experimental
observations, can be used as a research tool to reveal new
understanding of the biological systems.

Generic modeling is not about using sophisticated, as it may, mathematical
description to dress pre-existing understanding of complex biological
behavior. Rather, it means a cooperative approach, using existing 
biological
knowledge together with mathematical tools and synergetic point of view for
complex systems to reach a new understanding (which is reflected in the
constructed model) of the observed complex phenomena.

The generic models can yet be grouped into two main categories:
1. Discrete models such as the communicating walkers models of
Ben-Jacob \etal \cite{BSTCCV94a,BCSCV95,BCCVG97} and the bions model
of Kessler and Levine \cite{KL93,KLT97}.
In this approach, the microorganisms (bacteria in the first model and
amoebae in second) are represented by discrete, moving
entities (walkers and bions, respectively) which can consume
nutrients, reproduce, perform random or biased movement, and produce
or respond to chemicals. The time evolution of the chemicals is
described by reaction-diffusion equations.
2. Continuous or reaction-diffusion models \cite{PS78,Mackay78}. In
these models the microorganisms are represented via their 2D density,
and a reaction-diffusion equation of this density describes their
time evolution. This equation is coupled to the other
reaction-diffusion equations for the chemical fields. In the context
of branching growth, this idea has been pursued recently by Matsushita
\etal \cite{MWIRMSM98},
Kawasaki \etal \cite{KMMUS97} and Kitsunezaki \cite{Kitsunezaki97}. A
summary and critique of this approach is provided by Rafols
\cite{Rafols98}.

One of the important features of of the bacterial colonies
is the lubricant layer in which the bacteria swim.
A model for the colony should include this feature, directly or 
indirectly \cite{BCL98,GKCB98}.
Here we present a model which specifically includes the lubricant 
excreted by the bacteria.
The model follows the second approach of generic modeling.
We represent the various entities: the bacteria, the chemicals and 
the lubricant by continuous fields.

 \section{Experimental and Biological Background}

In figure 1a we show branching patterns of bacterial  
colonies. Each of these colonies is made up of about $10^{10}$ bacteria of
the type \Tname* (see \cite{BSST92,BTSA94} for first reference in the  
literature and \cite {TBG98} for identification). 
Each colony is  grown in a standard petri-dish ($8cm$ in diameter) on
a thin layer of agar (semi-solid  jelly). Figure 2a shows
that the branches of the colonies have well defined boundary, and
the bacteria are confined by this boundary. Figures 2b and 2c highlight
the constituents of the branches. Figure 2b shows that each branch is
a layer of fluid on the surface of the agar. Figure 2c shows the
bacteria, all of which are confined within this fluid.
The bacteria cannot move on the dry surface and cooperatively  they
produce a layer of lubrication fluid in which they swim.

Bacterial swimming is a random-walk-like movement, in which the  
bacteria propel themselves in nearly straight runs separated by brief  
tumbling. Swimming can be done only in sufficiently inviscid fluid. To
produce such fluid the bacteria secrete lubricant (wetting agents).
Other bacterial species produce known extracellular lubricants
(also known as surfactants, see \cite{MKNIHY92,MNZ93,DB97} and 
references therein).
These are various materials (various cyclic lipopeptides were identified) 
which draw water from the agar.  The composition and properties of 
the lubricant of \Tname is not known, but we will assume that higher 
concentration of lubricant 
is needed to extract water from a dryer agar, and that the lubricant
is slowly absorbed into the agar (or decomposes).

In order to move, reproduce and perform other metabolic activities, the
bacteria consume nutrients from the media, nutrients
which are given in limited supply. The growth of a colony is limited  
by the diffusion of nutrients towards the colony -- the bacterial  
reproduction rate that determines the growth rate of the colony is  
limited by the level of nutrients available for the cells.
If nutrient is deficient for a long enough period of time, the
bacteria may enter a pre-spore state,
i.e. begin the process of sporulation.
They stop normal activity -- like movement -- and use all their
internal reserves to metamorphose from an active volatile cell to a
spore -- sedentary durable ``seed''. The sporulating bacteria emit a
wide range of materials, some of which unique to the sporulating bacteria.
These emitted chemicals might be used by other bacteria as a signal
carrying information about the conditions at the location of the
pre-spores.

The patterns in figure 1a are arranged in a diagram according to two
control parameters: the initial concentration of nutrients (horizontal
axis, increasing from left to right), and the concentration of the
agar, or the dryness of the media (vertical axis, increasing from
bottom to top). The chirality of the colonies at the top row is due to
interaction between repulsive chemotaxis (see Sec. \ref{sec:chemo}) and 
the process of tumbling (see Refs. \cite{BCSCV95,BenJacob97} for details).
It will be ignored in this paper.
For high level of initial nutrients concentration (right column), the
patterns are compact, with wide branches. For intermediate levels of
nutrients, the lower the initial concentration is, the more ramified
and less ordered the patterns are. The patterns become fractal-like,
with fractal dimension decreasing for lower levels of nutrients. 
For the same nutrient level, higher agar concentration makes the
pattern less dense, with larger gaps between the branches.
All of the above phenomena could be expected from
our knowledge on patterning in non-living systems
\cite{KKL88,Langer89,BG90,BenJacob93,BenJacob97}.
Unlike what could have been expected, at the lowest concentration of
nutrient (leftmost column), the pattern are more ordered with a
well defined circular envelope.
This phenomenon demonstrates the complexity of the biological system,
and its explanation needs an additional biological feature --
chemotaxis signaling (see section \ref{sec:chemo}). Figure 3 demonstrates 
that in spite of this
complexity and the inherit noise in the system, the experiments are
controlled enough for the patterns to be reproducible.

\section{Construction of the Model}
\label{sec:construction}

The model includes four coupled fields projected on 2D.
One field describes the density of motile bacteria $b(\vec{x},t)$, 
the second describes the height of the lubricant layer in which the 
bacteria swim $l(\vec{x},t)$, third field describes the concentration 
of nutrients $n(\vec{x},t)$ and the fourth field is the density of 
stationary bacteria $s(\vec{x},t)$,
bacteria that enter the pre-spore state.

We first describe the dynamics of the bacteria and of the nutrient.
The two reaction-diffusion equations governing those fields have 
the following form:
\begin{eqnarray}
  \label{words}
  \partderiv{b}{t}&=&movement + reproduction - sporulation  \\
  \partderiv{n}{t}&=&\diffusion - consumption  \nonumber
\end{eqnarray}
The sporulation term refers to the transition of motile bacteria into 
the stationary state, i.e. the pre-spore state.
The nutrient diffusion is a simple diffusion process with a constant 
diffusion coefficient.
The rate of nutrient consumption is proportional to the rate of 
bacterial reproduction. 
For the reproduction term in (\ref{words}) we take:
\begin{equation}
\label{reproduction}
  reproduction=knb
\end{equation}
where $k$ is a constant rate. This is the usual form included in 
reaction-diffusion models, but see
section \ref{sec-new} for more detailed discussion.
The exact form of the sporulation term is not known. 
For simplicity we take the form:
\begin{equation}
  \label{mu_term}
  sporulation=\mu b
\end{equation}
where the rate $\mu$ is constant.

We now turn to the bacterial movement. In a uniform layer of liquid, 
bacterial swimming is a random walk with variable step length and can 
be approximated by diffusion. The layer of lubricant is not 
uniform, and its height affects the bacterial movement. An increase 
in the amount of lubricant decreases the friction between the 
bacteria and the agar surface. 
The term 'friction' is used here in a very loose manner
to represent the total effect of any force or process that slows down
the bacteria. It might include, for example, the drag which acts on a
body moving in shallow layer of viscous fluid. It might include the
probability that a flagellum will adhere or get tangled with the
polymers of the agar.
We suggest that the bacterial movement depends on the local lubricant 
height through a power law with the exponent $\gamma>0$:
\begin{equation}
  \label{bacterial_diffusion}
  movement=\nabla \cdot (D_b(l/l_M)^\gamma \nabla b)
\end{equation}
where $D_b$ is a constant with dimensions of a diffusion coefficient 
and $l_M$ is the height preferred by the bacteria (see below). 
$D_b$ is related to the fluid's viscosity and the dryness of the agar
might affect this viscosity. But as we argue below, this is probably 
not the main influence of the agar dryness.

Gathering the various terms gives the partial model:
\begin{eqnarray} 
  \label{eq3}
  \partderiv{b}{t}&=&\nabla \cdot (D_b(l/l_M)^\gamma \nabla b)+ kbn-\mu b
  \nonumber\\
  \partderiv{n}{t}&=&D_n\nabla^2n-\alpha kbn  \\
  \partderiv{s}{t}&=&\mu b \nonumber 
\end{eqnarray}
where $\alpha$ is a conversion factor, being the amount of nutrient consumed
for reproduction of a new bacterium.
The third equation in (\ref{eq3}) describes the stationary bacteria. 
Since they are immotile their dynamics include only a source term: 
their conversion from the motile state. 

The dynamics of the lubricating fluid are also governed by a reaction 
diffusion equation. There are two reaction terms: production by the 
bacteria and absorption into the agar.
The dynamics of the field are given by:
\begin{equation}
  \label{leq}
  \partderiv{l}{t}=-\nabla  \vec{J_l} + f_l(b,n,l) - \lambda l
\end{equation}
where $\vec{J_l}$ is the fluid flux 
(to be discussed), $f_l(b,n,l)$ is the fluid production term and 
$\lambda$ is the absorption rate of the fluid into the agar. 

We assume that the fluid production depends on the bacterial density.
As the production of lubricant probably demands substantial metabolic 
efforts, it should also depend on the nutrient's level. We take the 
simplest case where the production depends linearly on the concentrations 
of both the bacteria and the nutrients.
We presume that the bacteria produce lubricant only when it is needed, 
i.e. when its height is below a threshold height, denoted as $l_M$.
We therefore take the production term to be:
\begin{equation}
  f_l(b,n,l)=\Gamma bn(l_M-l)
\end{equation}
where $\Gamma$ is the production rate.

We turn to the flux of the lubricating fluid. 
The lubricating fluid flows by diffusion and by convection caused 
by bacterial motion.
A simple description of the convection is that as 
each bacterium moves, it drags along with it the fluid surrounding it.
\begin{equation}
  \vec{J_l}=-D_l(l/l_M)^\nu \nabla l + j\vec{J_b}
\end{equation}
where $D_l$ is a constant with dimensions of a diffusion coefficient,
$\vec{J_b}$ is the 
bacterial flux and $j$ is the amount of fluid dragged by each bacterium.
The diffusion term of the fluid depends on the height of 
the fluid to the power $ \nu >0$ (the nonlinearity in the diffusion 
of the lubricant, a very complex fluid, is motivated by 
hydrodynamics of simple fluids). The nonlinearity causes the fluid to 
have a sharp boundary at the front of the colony, as is observed in 
the bacterial colonies.
The equation for the lubricant field is:
\begin{equation} 
  \label{eql}
  \partderiv{l}{t}=\nabla \cdot ( D_l (l/l_M)^\nu \nabla l+
                    jD_b (l/l_M)^\gamma \nabla b )
  +\Gamma bn(l_M-l) - \lambda l 
\end{equation}
The functional form of the terms that we purposed are simple and plausible,
but they were not derived from basic physical principles. 
Therefore we cannot derive exact relations between the parameters of 
those terms and the physical properties of the agar substrate. 
However we can propose some  relations between the parameters and the agar.
In the experiments, the agar concentration is controlled. 
Higher agar concentration gives a drier and more solid substrate.
The lubricating fluid is composed of lubricant and water.
Increasing the agar concentration has several effects on the fluid and 
its dynamics.
Since the agar will be drier the absorption rate $\lambda$ will increase.
It will also make it more difficult for the lubricant to extract 
water from the agar substrate.
Therefore the production rate $\Gamma$ should decrease. 
More lubricant will be required to extract water so the composition 
of the fluid will change. The lubricant concentration will increase, 
making the fluid more viscous,
thus slowing its flow. Therefore the diffusion coefficient $D_l$ 
should decrease.
A viscous fluid will also slow the bacterial movement, so the bacterial 
diffusion coefficient $D_b$ should also decrease.

Equation (\ref{eql}) together with equations (\ref{eq3}) form our model.
Before further studies of the model
we reduce the number of parameters by using dimensionless units. 
We define the new variables:
\begin{equation}
\begin{array}{lcr}
\label{transformation}
  t' = t \mu, &
  \vec{x}' = \vec{x} \sqrt{\mu/D_n},  \\ 
  b' = b k\alpha/\mu, & \nonumber 
  n' = n k/\mu, &
  l' = l/l_M
\end{array}
\end{equation}
With the same units we define:
\begin{equation}
\begin{array}{lcr} 
\label{newparam}
  D_b'=D_b/D_n, &
  D_l'=D_l/D_n, \\
  \Gamma'=\Gamma \mu/k^2\alpha, & \nonumber 
  \lambda'=\lambda /\mu,  &
  j'=j\mu/k\alpha l_M 
\end{array}
\end{equation}
Using these variables in (\ref{eq3}) and (\ref{eql})
and omitting the primes we get:
\begin{eqnarray} 
  \label{model}
  \partderiv{b}{t}&=&\nabla \cdot (D_bl^\gamma \nabla b)+ bn-b \nonumber \\
  \partderiv{n}{t}&=&\nabla^2n-bn  \\
  \partderiv{l}{t}&=&\nabla \cdot (D_l l^\nu \nabla l+jD_b l^\gamma 
  \nabla b) +\Gamma bn(1-l) - \lambda l \nonumber \\
  \partderiv{s}{t}&=& b \nonumber 
\end{eqnarray}

\section{Results of Numeric Simulation}
\label{sec:simulation}

Figures 4-10 show results of numerical simulations of the model.
The figures display the sum of the active and stationary 
bacterial densities, $b+s$.
The simulation were done with an explicit method.
To reduce the implicit lattice anisotropy, 
a quenched noise was introduced into the diffusion operators.
For the initial conditions, we set $n$ to have uniform distribution 
of level $n_0$, $b$ to be zero everywhere but in the center, and the 
other fields to be zero everywhere.

In figure \ref{fig-fields} we show the fields $s$, $n$, $b$ and $l$. 
The nutrient field $n$ was consumed by bacteria 
down to a level of $n \sim 1$ ($\mu/k$ in the original units) in 
the area covered by the colony
(the nutrient was not completely depleted due to the functional form 
of the bacterial sporulation term, the same as in \cite{Kitsunezaki97} 
and unlike \cite{KMMUS97,MWIRMSM98}).
As the nutrient diffuses faster than the other fields it also decreased 
in the area between the colony branches. 
The field of the motile bacteria $b$ and the lubricant
field $l$ are overlapping. The motile bacteria are confined
to the area covered by lubricant.
The fronts of both fields have compact support as
 figure \ref{fig-fronts} shows.

In experiments of the bacterial colonies there are two control parameters:
the agar concentration and the initial nutrient concentration.
First we examined the effect of changing the latter. 
As figure \ref{fig-nutrient} shows, we obtained 
a dense circular colony when $n_0$ was large, a branched pattern when 
we decreased $n_0$
and a DLA-like pattern when $n_0$ was close to $1$. 
Similar effects of decreasing the initial nutrient level appear in other 
reaction-diffusion models \cite{MWIRMSM98,KMMUS97,Kitsunezaki97,GKCB98}.

Changing the agar concentration affects the dynamics of the lubricant 
fluid. We suggested in the previous section that a higher agar 
concentration relates to a larger absorption rate $\lambda$ and to lower
production rate $\Gamma$ and diffusion coefficients $D_l$ and $D_b$.
It is not a-priori clear what is the exact dependence of each parameter on
the agar concentration. We use the model to investigate this question.
In figure \ref{fig-lambda} we show  patterns obtained with different values 
of the parameters $\Gamma$ and $\lambda$. As we expected, 
increasing $\lambda$ or decreasing $\Gamma$ produced a more ramified 
pattern, similar to the effect of higher agar concentrations on the 
patterns of bacterial colonies (figure 1a).
The value of the diffusion coefficient of the lubricating fluid $D_l$
has less influence on the colony pattern, as can be seen in 
figure \ref{fig-Dl}.
In contrast, decreasing the bacterial diffusion coefficient $D_b$ produces
a ramified pattern, as figure \ref{fig-Db} shows.

In most of the figures we took $\gamma=1$ and $\nu=1$.
There is no a-priori reason to take these values, 
and in figure \ref{fig-exponents} 
we show the effect of other values on the growth.
While the patterns are different, we found that these changes can be 
compensated by other parameters and varying $\gamma$ and $\nu$ has no 
qualitative effect on the conclusions. 

\section{Chemotaxis}
\label{sec:chemo}

The model so far reproduced most of the features of the experimental 
results displayed in figure 1a, but does not reproduce the transition to 
ordered patterns at the lowest nutrient concentration.
We will now extend the lubricant model to test for its
success in describing this phenomenon.
Ben-Jacob \etal suggested that this transition is due to
chemotaxis and chemotactic signaling \cite{BCC96,CCB96,BC97,BenJacob97}.
Chemotaxis means changes in the movement of the cell in response to a
gradient of certain chemical field
\cite{Adler69,BP77,Lacki81,Berg93}. The movement is biased along the
gradient either in the gradient direction or in the opposite
direction. Usually chemotactic response means a response to an
externally produced field, like in the case of chemotaxis towards
food. However, the chemotactic response can be also to a field
produced directly or indirectly by the bacterial cells. We will refer
to this case as chemotactic signaling.

We incorporate the effect of chemotaxis by
introducing a bacterial flux $\vec{J}_{chem}$ due to chemotaxis. 
The general form is:
\begin{equation}
\vec{J}_{chem}\equiv \zeta (l)b\chi (r)\nabla r
\label{j_chem}
\end{equation}
where $r$ is the chemotactic chemical.
$\chi (r)\nabla r$ is the gradient sensed by the bacterium (with $\chi
(r)$ having the units of 1 over chemical's concentration). $\chi (r)$
is taken to be the ``receptor law'', i.e. $\chi (r)=K_r/(K_r+r)^2$ 
\cite{Murray89}.
$K_r$ is a constant that determines the concentration range of the chemical
for which the chemotaxis is effective.
$\zeta (l)$ (having the same units as a diffusion coefficient)
is the bacterial response to the sensed gradient
(i.e. the effect on the bacterial movement). 
In our  model the bacterial diffusion coefficient is $D_bl^\nu$, and the
bacterial response to chemotaxis is $\zeta (l)=\zeta _0\left(
D_bl^\nu \right)$. $\zeta _0$ is a constant,
positive for attractive chemotaxis and negative for repulsive
chemotaxis.

{\em Amplification of diffusive Instability Due to Nutrients
Chemotaxis:}
In non-living systems, more ramified patterns (lower fractal
dimension) are observed for lower growth velocity. Based on growth
velocity as function of nutrient level and based on growth dynamics,
Ben-Jacob \etal \cite {BSTCCV94a} concluded that in the case of
bacterial colonies there is a need for mechanism that can both
increase the growth velocity and maintain, or even decrease, the
fractal dimension. They suggested food chemotaxis to be the required
mechanism. It provides an outward drift to the cellular movements;
thus, it should increase the rate of envelope propagation. At the
same time, being a response to an external field it should also
amplify the basic diffusion instability of the nutrient field. Hence,
it can support faster growth velocity together with a ramified
pattern of low fractal dimension.
The bacterial flux due to nutrient chemotaxis is:
\begin{equation}
\vec{J}_{nutrient}\equiv \zeta_n D_bl^\nu \chi_n  
                         b\frac{K_n}{(K_n+n)^2}\nabla n
\end{equation}

In figure \ref{fig-foodchemo} it is
shown that as expected, the inclusion of food chemotaxis 
led to a considerable increase of the growth velocity without
significant change in the fractal dimension of the pattern.

{\em Repulsive chemotactic signaling: }
We focus now on the formation of the fine radial branching patterns
at low nutrient levels. From the study of non-living systems, it is
known that in the same manner that an external diffusion field leads
to the diffusion instability, an internal diffusion field will
stabilize the growth. It is natural to assume that some sort of
chemotactic agent produces such a field. To regulate the organization
of the branches, it must be a long-range signal. To result in radial
branches it must be a repulsive chemical produced by bacteria at the
inner parts of the colony. The most probable candidates are the
bacteria entering a pre-spore state, 
which were referred to as the stationary bacteria $s$.

As stated above, bacteria may enter a pre-spore state upon starvation.
In this process they
emit a wide range of waste materials, some of which unique to
the sporulating bacteria. These emitted chemicals might be used by other
bacteria as a signal carrying information about the conditions at the
location of the pre-spores. Ben-Jacob \etal \cite
{BSTCCV94a,BSTCCV94b,CCB96} suggested that such materials are
repelling the bacteria ('repulsive chemotactic signaling') as if they
escape a dangerous location.

The equation describing the dynamics of the chemorepellent contains
terms for diffusion, production by pre-spores, decomposition by
active bacteria and spontaneous decomposition:
\begin{equation}
\partderiv{r}{t}=D_r{\nabla ^2 r+\Gamma }_r s-\Omega _r b r- \lambda_r r
\label{r-eqn}
\end{equation}
where $D_r$ is the diffusion coefficient of the chemorepellent,
$\Gamma _r$ is the emission rate of repellent by pre-spores, $\Omega
_r$ is the decomposition rate of the repellent by active bacteria,
and $\lambda _r$ is the rate of self decomposition of the repellent.
The bacterial flux due to repulsive chemotaxis is:
\begin{equation}
\vec{J}_{repulsive}\equiv \zeta_r D_bl^\nu \chi_r  
                         b\frac{K_r}{(K_r+r)^2}\nabla r
\end{equation}
where $\zeta_r<0$ (repulsive chemotaxis).

In figure \ref{fig-chemotaxis} the effect of repulsive chemotactic 
signaling is shown. In the presence of repulsive chemotaxis the patterns
have a smooth circular envelope, while the
branches are thinner and radially oriented.

\section{Another look on modeling bacterial growth}
\label{sec-new}

The model presented in section (\ref{sec:construction}) goes along the 
lines of existing models \cite{Kitsunezaki97,MWIRMSM98} in interpreting 
the bacterial growth terms (reproduction and sporulation).
There are, however, some biological consideration that may have been 
overlooked. In section \ref{sec:construction} we termed $knb$ 
as 'reproduction' and $\mu b$ as 'sporulation'. 
But looking at them term by term unravel discrepancies.
The so called ``sporulation'' term in the equations represents, 
according to the usual interpretation, a constant probability per 
time unit for a bacterium to enter the pre-spore state.
This is incompatible with our biological knowledge. A nutritional 
stress is one of 
the pre-requirements for starting the process of sporulation. As long
as there is enough food for the bacteria to reproduce, there is 
no nutritional stress. In the model it means that as long as
the (local) bacterial density increases (i.e. as long as the sum
of all the reaction terms is positive) there can be no sporulation.

This leads to a new interpretation of the reaction terms;
the resources that the bacteria consume from the nutrient 
(energy and materials)
are utilized for two main processes: to sustain life and to reproduce.
Therefore the reproduction rate should be proportional to the 
nutrient consumed less the amount required for life sustaining activities.
We assume that the latter is required by each bacterium at a constant rate,
independent of the nutrient level or the bacterial density. We denote 
that rate by $\mu$, which previously denoted
the sporulation rate. Then we have instead of (\ref{reproduction}), 
with $g(b,n)$ denoting the nutrient consumption rate:
\begin{eqnarray}
\label{reproduction2}
  reproduction &=& g(n,b) - \mu b \\
  sporulation &=&
\left\{ \begin{array}{ll}
0 & \mbox{~~~if } reproduction > 0 \\
-reproduction & \mbox{~~~if } reproduction < 0 
\end{array} \right.
\hspace{0.2cm}= \max \left(\mu b-g(n,b),0 \right)
\end{eqnarray}

So far we have not changed much the model. All the terms involved in 
the dynamics remain the same,
but we give them new interpretations.
Only the dynamics of the field $s$ changed. Since the dynamics of the 
other fields are decoupled from the field $s$, the modified model 
is effectively identical to the previous model. 

As we changed the meaning of the bacterial growth terms, 
we should reconsider their functional form,
focusing our attention on the nutrient consumption term, $g(n,b)$.
The term $g(n,b)=knb$ was taken to be the limit of growth at 
low nutrient concentration, but it is not a-priori evident. 
We suggest that $g$ should  have three regimes corresponding to different 
limits of the nutrient concentration and  bacterial density.
\begin{enumerate}
\item There is a maximal growth rate of the bacteria, 
along with a maximal rate of nutrient consumption, even for optimal 
conditions. 
We denote the maximal consumption rate per bacterium by $\Omega_n$.
\item When conditions are not optimal and the nutrient concentration is low, 
it might be the limiting factor. 
Nutrient consumption is then diffusion-limited and its rate is proportional
to $D_n n$ per bacterium, as $D_n$ defines the effective area from which a
bacterium consumes food (assuming no other bacteria
interfere with this process).
\item When the nutrient concentration is low and the bacterial density 
is high, there is competition 
between the bacteria, and the amount of nutrient available for each 
bacterium is proportional to $n/b$. 
\end{enumerate}
As can be seen in figure 2c, the density of bacteria can be quite high 
and so the third limiting
behavior cannot be ignored. The ratio between the nutrient diffusion length
and the distance between bacteria $b^{-1/2}$ determines which is 
the appropriate regime. We take the nutrient consumption rate to be 
the minimum of the three expressions: 
\begin{equation}
\label{gbn}
   g(b,n) = b \; \min(\Omega_n, k n, en/b)
\end{equation}
where $k=D_n/\alpha$ and $e$ is a constant rate (the min function could 
have been replaced by a smooth function with the same limits, but we 
will put it in this way for clarity). We note that the Michaelis-Menten
law \cite{Murray89} of $\frac k{1+gn}nb$ as the first  two
expressions as limits, but does not incorporate  the third one.
The new model in dimensionless form is:
\begin{eqnarray} 
  \label{modelnew}
  \partderiv{b}{t}&=&\nabla \cdot (D_bl^\gamma \nabla b)+ g(n,b)-b 
                      \nonumber  \\
  \partderiv{n}{t}&=&\nabla^2n- g(n,b) \nonumber  \\
  \partderiv{l}{t}&=&\nabla \cdot (D_l l^\nu \nabla l+jD_b l^\gamma 
    \nabla b) +\Gamma bn(1-l) - \lambda l \\
  \partderiv{s}{t}&=&  \max(b-g(n,b),0) \nonumber 
\end{eqnarray}
where $g(n,b) =  b \; \min(\frac{\Omega_n}{\mu}, n,
\frac{e \alpha}{\mu}\frac n b)$
The same transformation of variables (\ref{transformation}) 
and (\ref{newparam}) were employed to obtain the dimensionless model.

Numerical simulations of the above model show (figure \ref{fig-new}) 
no qualitative differences from the 
model presented in section \ref{sec:simulation}. It is not the 
mathematical model which is important in this respect, but our 
understanding of the biological system. This emphasizes the differences
between the 'generic modeling approach' and the approach driving 
the ``reminiscence syndrome''

\section{Conclusion}

We first briefly reviewed experimental observations of branching patterns 
in bacteria of the species {\it Paenibacillus }. 
Both colonial patterns and 
optical microscope observations of the bacteria dynamics were presented.

Our goal in this manuscript was to test a new reaction-diffusion model
which includes time evolution equation of a lubricant. From a comparison
of the model simulation and experimental observations we conclude that 
when a specific bacterial strain is considered, such comparison is 
not sufficient to
tell us if indeed the right biological features are included in the model.

For more critique test of the models, additional aspects of the growth
(such as functional dependence of the colonial growth velocity on
growth conditions, branches size and width distributions etc.) have to
be compared with the model predictions.
One should also compare the theory with more involved experimental
tests, such as the effect of imposed anisotropy, competition between
neighboring colonies, and expression of mutants (emergence of
sectors) in expanding colonies. 

Our conclusion from the study of bacterial branching growth is that
the minimal features of diffusion, food consumption,
reproduction and
inactivation are not sufficient to explain the complete picture of the
observed phenomena. We believe that additional mechanisms must be 
introduced, and propose chemotactic signaling as plausible one.

This work has dealt with a continuous model. Such models are not 
preferable to discrete ones.
Each has its advantages and disadvantages. The discrete walkers model, for
example, enables us to include the valuable feature of internal degrees
of freedom, but is computationally limited in the number of walkers
that can be simulated, and thus its scaling to the real problem is
somewhat problematic.
The best strategy is to employ in parallel both the reaction-diffusion
and the walkers approaches.


\newpage
\begin{figure}[th]
\caption{
a) A diagram of colonial branching patterns of the bacteria
\protect{\Tname*}.
The dot at the center of each colony is the initial inoculum.
The horizontal axis of the diagram is initial nutrient
concentration. From left to right (in units of $gram/liter$) it is
0.1, 0.5, 1.0, 2.0 and 3.0 .
The vertical axis is agar  concentration. From bottom to top (in units
of $10 gram/liter$) it is 1.5, 2 and 2.5 .
As the level of initial nutrient concentration is decreased the
patterns become less organized with fewer branches. However, at the
lowest concentration, of $0.1g/l$, the pattern is ordered with
circular envelope.
b) An isolated example of branching pattern at $0.5 g/l$ pepton
and $1.75\%$ agar concentration.
}
\end{figure}
\begin{figure}[th]
\caption{
Closer look on branches of a colony.
a) $\times20$ magnification shows the sharp boundaries of the branches.
The width of the boundary is in the order of micron.
b) Numarsky (polarized light) microscopy shows the hight of the
branches and their envelope. What is actually seen is the layer of
lubrication fluid, not the bacteria.
c) $\times50$ magnification shows the bacteria inside a branch. Each
bar is a single bacterium. There are no bacteria outside the branch.
}
\end{figure}
\begin{figure}[th]
\caption{
Demonstration of the reproducibility of the colonial patterns.
Four colonies in the same plate (from four inocula). 
}
\end{figure}
\begin{figure}[th]
\caption []{
A representative branching pattern produced by the lubricating bacteria 
model. The image shows the four fields of the model.
}
\label{fig-fields}
\end{figure}

\begin{figure}[th]
\caption []{
  Profile of the fronts of the bacterial field (solid line)
  and of the lubricating field (dashed line) from the 2D model.
  The fields propagate to the right.
  Both fronts have compact support.
}
\label{fig-fronts}
\end{figure}

\begin{figure}[th]
\caption []{
Effect of varying the initial nutrient concentration $n_0$ on colony 
pattern. The minimal value of $n_0$ to support growth is $1$.
We include the time $t$ it took for the colony to grow.\\
a. $n_0=1.1, t=38089$, DLA-like pattern. \\
b. $n_0=1.3, t=11410$. \\
c. $n_0=1.7, t=3671$, branched pattern. \\
d. $n_0=3, t=721$, dense branches. \\
e. $n_0=5, t=216$, a disk.\\
The other parameters are: $D_b=D_l=0.5, \Gamma=0.6, \lambda=0.1, 
j=0.01, \gamma=1, \nu=2$.
}
\label{fig-nutrient}
\end{figure}

\begin{figure}[th]
\caption []{
Effect of varying $\lambda$, the fluid absorption rate, on colony pattern. 
The fluid production rate $\Gamma$ is $1$ in the upper row,
and $0.3$ in the lower row.
In both rows $\lambda$ increases from left to right:
$\lambda=0.03$ (left), $\lambda=0.1$ (center), $\lambda=1$ (right)
The patterns become more ramified as $\lambda$ increases. 
Here and in figures \ref{fig-Dl}, \ref{fig-Db} decreasing $\Gamma$ also
produces a more ramified pattern.
The other parameters are:
$D_b=D_l=1, \gamma=\nu=1, n_0=1.5, j=0.01$
}
\label{fig-lambda}
\end{figure}

\begin{figure}[th]
\caption []{
Effect of varying $D_l$, the fluid diffusion coefficient,
on colony pattern. 
Fig. (a) and (d) are the same as Fig. \ref{fig-lambda} (a) and (d).
In both rows $D_l$ decreases from left to right:
$D_l=1$ (left), $D_l=0.1$ (center), $D_l=0.01$ (right).
The patterns do not change significantly.
}
\label{fig-Dl}
\end{figure}

\begin{figure}[th]
\caption []{
Effect of varying $D_b$, the bacterial diffusion coefficient,
on colony pattern. 
Fig. (a) and (d) are the same as Fig. \ref{fig-lambda} (a) and (d).
In both rows $D_b$ decreases from left to right:
$D_l=1$ (left), $D_l=0.1$ (center), $D_l=0.05$ (right).
The patterns become more ramified as $D_b$ decreases. 
}
\label{fig-Db}
\end{figure}

\begin{figure}[th]
\caption []{
Effect of changing the exponents $\gamma$ and $\nu$.
We also print the growth time of the colony $t$.
The parameters of the figures are:\\
a. $\gamma=1, \nu=1, t=1249$. \\ 
b. $\gamma=1, \nu=2, t=2069$. \\
c. $\gamma=2, \nu=1, t=4451$. \\
d. $\gamma=2, \nu=2, t=10066$. \\
The other parameters are: $D_b=D_l=1, n_0=1.5, \Gamma=1, 
\lambda=0.03, j=0.01$.
}
\label{fig-exponents}
\end{figure}

\begin{figure}[th]
\caption []{
The effect of food chemotaxis on growth.
The four patterns differ in the values of $\zeta_n$, 
the response to the sensed gradient of the nutrient.\\
We also print the growth time of the colony $t$.\\
a. $\zeta_n=0, t=5101$ (no chemotaxis). \\
b. $\zeta_n=10, t=4957$. \\
c. $\zeta_n=30, t=3843$. \\
d. $\zeta_n=100, t=2053$. \\
The other parameters are: $D_b=D_l=1, \gamma=\nu=1, n_0=1.5, 
\Gamma=0.3, \lambda=0.1, j=0$.
}
\label{fig-foodchemo}
\end{figure}

\begin{figure}[th]
\caption []{
The effect of repulsive chemotaxis on growth.
The four patterns differ in the values of $\zeta_r$, 
the response to the sensed gradient of the chemorepellent.\\
a. $\zeta_r=0$ (no chemotaxis). \\
b. $\zeta_r=-20$. \\
c. $\zeta_r=-50$. \\
d. $\zeta_r=-100$. \\
The other parameters are: $D_b=D_l=1, \gamma=\nu=1, 
n_0=1.3, \Gamma=0.3, \lambda=0.1, j=0$.
}
\label{fig-chemotaxis}
\end{figure}

\begin{figure}[th]
\caption []{
Pattern produced by the modified model of section \ref{sec-new}
}
\label{fig-new}
\end{figure}
\end{document}